# Minimum-Cost Synthetic Genome Planning: An Algorithmic Framework

Michail Patsakis[1], Ioannis Mouratidis[1], Ilias Georgakopoulos-Soares[1,*]

[1] Division of Pharmacology and Toxicology, College of Pharmacy, The University of Texas at Austin, Dell Paediatric Research Institute, Austin, TX, USA.
[*] Corresponding authors: ilias@austin.utexas.edu

**Abstract**

As synthetic genomics scales toward the construction of increasingly larger genomes, computational strategies are needed to address technical feasibility. We introduce an algorithmic framework for the Minimum-Cost Synthetic Genome Planning problem, aiming to identify the most cost-effective strategy to assemble a target genome from a source genome through a combination of reuse, synthesis, and join operations. By comparing dynamic programming and greedy heuristic strategies under diverse cost regimes, we demonstrate how algorithmic choices influence the cost-efficiency of large-scale genome construction. In parallel, solving the Minimum-Cost Synthetic Genome Planning problem can help us better understand genome architecture and evolution. We applied our framework in case studies on viral genomes, including SARS-CoV-2, to examine how source-target genome similarity shapes construction costs. Our analyses revealed that conserved regions such as ORF1ab can be reconstructed cost-effectively from related templates, while highly variable regions such as the S (spike) gene are more reliant on DNA synthesis, highlighting the biological and economic trade-offs of genome design.

## Introduction

Synthetic genomics has expanded rapidly in recent years, offering unprecedented opportunities to understand genome function and engineer organisms with novel capabilities (Coradini, Hull, and Ehrenreich 2020; James et al. 2025). Major milestones include the synthesis of a functional poliovirus genome of approximately 7,500 base pairs in length (Cello, Paul, and Wimmer 2002), followed by the synthesis of the first self-replicating artificial cell, *Mycoplasma capricolum* (Gibson et al. 2010) and the design and chemical synthesis of of the *S. cerevisiae* chromosomes (Schindler, Walker, and Cai 2024).

Although DNA sequencing costs have declined precipitously, the cost of de novo DNA synthesis remains comparatively high (Shendure et al. 2017; Hoose et al. 2023). Additionally, de novo DNA synthesis is currently restricted in terms of sequence length and fidelity, requiring the assembly of fragments through different techniques such as Golden Gate assembly, Gibson assembly, and DNA assembly in vitro or in vivo (Ma et al. 2024; Gibson et al. 2010; Zhang et al. 2020). DNA reuse of existing DNA sequences can decrease costs and improve the generation of long sequences by minimizing the need for de novo synthesis.

To navigate these complexities, computational tools have been developed to optimize assembly plans. Software such as Raven (Appleton et al. 2014) and DNALD (Blakes et al. 2014) provide heuristic solutions, particularly for the complex task of assembling DNA libraries where many target molecules are constructed by reusing shared intermediate parts. This reliance on heuristics is often a practical necessity, as the underlying combinatorial problems can be computationally intractable; for instance, a simplified version of library assembly has been proven to be NP-hard (Blakes et al. 2014), making guaranteed optimal solutions infeasible for large-scale problems.

Concurrently, foundational work in computational biology has focused on modeling the evolutionary distance between genomes. These models employ a set of "moves" intended to represent large-scale mutational events, including operations such as inversions, translocations and the more general double-cut-and-join (DCJ) (Yancopoulos, Attie, and Friedberg 2005; Ravi and Kececioglu 1995). These advances have been complemented by algorithmic studies on genome rearrangement distances, breakpoint graphs, and their applications to phylogenetic inference (Alekseyev and Pevzner 2009; Simonaitis, Chateau, and Swenson 2018; Pevzner and Tesler 2003). However, while powerful for studying evolution, this set of operations does not reflect the physical or economic realities of building a genome in a laboratory.

Here, we define and solve the Minimum-Cost Synthetic Genome Planning problem from a theoretical standpoint. We introduce a novel algorithmic framework that moves beyond the process-based metrics of library assembly and the evolutionary models of genome rearrangement. Instead, we utilize a set of operations (reuse, synthesis, and join) that directly correspond to laboratory procedures and incorporate a weighted cost system reflecting their distinct expenses. Our goal is to determine the most cost-effective strategy to construct a target genome by optimally partitioning it into blocks that are either replicated from an existing source genome or synthesized de novo. This work bridges the gap between classic rearrangement theory and the practical demands of genomic engineering and provides a novel lens through which to measure evolutionary distance.

## Methods

### Formal problem definition

Let the source genome be represented as a string $G_S$ and the target genome as a string $G_T$, both composed of characters from the nucleotide alphabet $\Sigma = \{A, C, G, T\}$. Our objective is to construct $G_T$ by concatenating a sequence of non-overlapping blocks, $b_1, b_2, \ldots, b_k$, such that their concatenation equals $G_T$. The cost of this construction is determined by the method used to acquire and assemble each block.

### Genomic operations and cost model

We define a set of operations based on a workflow inspired by Golden Gate assembly (Engler et al., 2009), where multiple DNA fragments are joined in a one-pot reaction.

A Reuse operation, corresponding to PCR amplification, involves identifying a block $b_i$ that exists as a substring within the source genome $G_S$. This operation has a low, fixed cost, denoted $C_{reuse}$, which accounts for primers and reagents, regardless of the block's length.

A Synthesis operation involves the de novo chemical synthesis of a block $B_k$. The cost of this operation, $C_{synth}$, is a function of the block's length $w$, reflecting that longer fragments are significantly more expensive to synthesize. We model this as a linear function $C_{synth}(w) = w \cdot c_s$, where $c_s$ is a constant representing the cost per base.

A Join operation represents the cost of assembling two adjacent blocks. In a Golden Gate context, this involves designing compatible overhangs and performing a ligation reaction. We abstract this into a fixed cost, $C_{join}$, which is incurred for each junction created between fragments. A construction of $k$ blocks will therefore require $k-1$ join operations.

### The Minimum-Cost Transformation Problem

Given $G_S$ and $G_T$, and the cost functions $C_{reuse}$, $C_{synth}(L)$, and $C_{join}$, the problem is to find a partition of $G_T$ into blocks $b_1, b_2, \ldots, b_k$ that minimizes the total construction cost, defined as:

$$TotalCost = \sum_{i=1}^{k} AcquisitionCost(b_i) + (k-1)C_{join}$$

where the $AcquisitionCost$ for each block is either $C_{reuse}$ or $C_{synth}(|b_i|)$.

### Algorithms

To solve the Minimum-Cost Synthetic Genome Planning Problem, we developed and compared three distinct algorithmic strategies. The first is a dynamic programming approach that guarantees an optimal solution, while the other two are greedy heuristics designed to find fast, approximate solutions based on different local optimization criteria.

**Dynamic programming algorithm for optimal planning**
To find the provably optimal solution, we employ dynamic programming. Let $T$ be the target genome of length $N$. We define an array, $DP$, where $DP[i]$ stores the minimum possible cost to construct the prefix of the target genome of length $i$, i.e., $T[1..i]$. The goal is to compute $DP[N]$.

The base case is $DP[0] = 0$, representing zero cost to construct an empty prefix. The recurrence relation to compute $DP[i]$ for $i > 0$ is defined as follows:

$$DP[i] = \min_{1 \leq w \leq W} \left( DP[i-w] + \text{Cost}(T[i-w+1..i]) + C_{join} \right)$$

where $W$ is the maximum allowed block length, a parameter that reflects the practical upper limit on the length of a DNA fragment that can be reliably synthesized or amplified. The term $\text{Cost}(T[i-w+1..i])$ is the acquisition cost of the block of length $w$ ending at position $i$. This cost is determined by querying the source genome $S$. If the block $T[i-w+1..i]$ is found in $S$, its cost is $C_{reuse}$. Otherwise, its cost is $C_{synth}(w)$.

The $C_{join}$ cost is omitted for the very first block (when $i - w = 0$). By iterating through all possible last blocks for each position $i$, this algorithm explores the entire solution space and guarantees that $DP[N]$ holds the global minimum construction cost.

**Replication-first greedy algorithm**
As a baseline for comparison, we implemented a greedy algorithm that prioritizes the reuse of existing genetic material. This "Replication-First" strategy iterates through the target genome from start to finish. Starting at position $i = 0$, it attempts to make a locally optimal choice by searching for the longest possible block starting at $i$ (up to length $W$) that exists in the source genome $S$.

If such a replicable block of length $w$ is found, the algorithm immediately selects it, adds $C_{reuse}$ and $C_{join}$ to the total cost (omitting $C_{join}$ if $i = 0$), and advances the position by $w$. If no replicable block of any length (from $W$ down to 1) is found starting at position $i$, the algorithm defaults to its only remaining option: synthesizing a single base. In this case, it adds the cost $C_{synth}(1)$ and advances the position by one. This process repeats until the entire target genome is constructed.

**Max-block greedy algorithm**
We designed a second greedy heuristic to explore an alternative strategy focused on minimizing the number of join operations. This "Max-Block" algorithm always attempts to construct the target genome using the largest possible fragments. At each position $i$, it invariably considers the block of length $w = \min(W, N - i)$.
It then makes a local cost decision for this max-sized block. It first checks if the block exists in the source genome $S$. If it does, the algorithm chooses the cheaper option between replicating it (cost $C_{reuse}$) and synthesizing it (cost $C_{synth}(w)$). If the block does not exist in the source, the algorithm has no choice but to synthesize it at cost $C_{synth}(w)$.

After selecting the block and adding the appropriate acquisition and join costs, the position is advanced by $w$. This strategy aggressively reduces the number of $C_{join}$ costs but may be forced into expensive synthesis operations that the Replication-First or DP algorithms would avoid.

**The break-even length**
To rigorously evaluate our algorithms under diverse economic trade-offs, we introduced a unified metric called the Break-Even Length, denoted as $W_{BE}$. This value represents the specific k-mer length at which the cost to synthesize a DNA block becomes more favorable than the fixed cost to replicate it. This trade-off is captured by the inequality $C_{synth}(w) \leq C_{reuse}$.

Given our linear cost model where $C_{synth}(w) = w \cdot c_s$, this inequality becomes $w \cdot c_s \leq C_{reuse}$. By solving for $w$, we can identify the range of lengths where synthesis is the cheaper acquisition method. The break-even point occurs at $W_{BE} = C_{reuse}/c_s$. For any block shorter than $W_{BE}$, synthesis is preferred, while for any block longer than $W_{BE}$ replication is favored. By systematically varying the $W_{BE}$ value (by adjusting $C_{reuse}$ and $c_s$ while holding $C_{join}$ constant), we can efficiently explore the entire spectrum of economic pressures and test the robustness of each algorithm's decision-making strategy.

**Genomic analysis of viral genomes**
To demonstrate the utility of our framework, we performed three computational case studies using viral genomes. These analyses were designed to (1) empirically evaluate the performance of the greedy heuristics against the optimal DP algorithm, (2) quantify the relationship between source-target genome similarity and construction cost, and (3) investigate whether cost profiles can reveal features of genome architecture.

**Analysis 1: Algorithm Performance Comparison**
To conduct a robust comparison of the Dynamic Programming (DP), Replication-First Greedy, and Max-Block Greedy algorithms, we performed a leave-one-out cross-validation using a diverse set of 86 viral genomes (Supplementary Table S1). For each of the 86 viruses in the dataset, it was designated as the target genome. The corresponding source genome was constructed by concatenating the remaining 85 viruses, from which a single FM-Index was built. We then calculated the construction cost for each of the three planners. This entire process was repeated for a range of eight distinct "Break-Even Length" ($W_{BE}$) values, from 5 to 25, to assess performance across different economic trade-offs between replication and synthesis. The maximum block length (W) was held constant at 200 bp.

**Analysis 2: Impact of Source Genome Similarity on Construction Cost**
To investigate how evolutionary distance impacts economic feasibility, we calculated the optimal construction cost of a fixed target genome from a panel of source genomes with varying degrees of similarity. The reference genome of SARS-CoV-2 ($NC045512.2$) was selected as the target. A curated set of 12 source genomes from the Coronaviridae family was used, chosen to provide a range of genetic similarities (Supplementary Table S2). For each of the 12 source genomes, a dedicated FM-Index was built. The optimal construction cost of the target was then calculated using this index. This procedure was repeated for four distinct economic scenarios: "Replication-Dominant," "Synthesis-Dominant," "High-Join-Cost," and "Balanced." The similarity between each source and the target was quantified post-hoc using the Average Nucleotide Identity (ANI), estimated with the MASH software.

**Analysis 3: Genome Architecture Cost Profiling**
To test the hypothesis that our cost model can identify functionally conserved and variable genomic regions, we generated a "cost profile" of the SARS-CoV-2 genome ($NC045512.2$). This was accomplished by modifying our DP planner to output the entire cost array, $DP[0...N]$. The local construction cost for a given region was then calculated using a 500 bp sliding window. This analysis was performed in two distinct experiments. First, a single cost profile was generated using the closely related Bat coronavirus RaTG13 ($MN996532.2$) as the source. Second, to obtain a consensus view, cost profiles were generated from a curated set of eight different coronavirus source genomes (Supplementary Table S3). These individual profiles were then used to calculate a mean cost profile and its standard deviation across the target genome (Figure 2B). Both experiments were run using the "Balanced" cost parameters ($C_{reuse} = 5$, $C_{join} = 1.5$, $C_{synth} = 0.2$) with a maximum block length $W = 100$

## Results

**Experimental design:**

**Analysis of algorithmic performance**
We compared the performance of the two greedy heuristics relative to the optimal Dynamic Programming (DP) planner. The results demonstrate that while the DP algorithm consistently finds the optimal solution, the performance of the greedy algorithms is highly sensitive to the economic conditions defined by the $W_{BE}$ (**Figure 1a**). The Replication-First Greedy algorithm, which is designed to prioritize the reuse of genetic material, performs well when $W_{BE}$ is low, as this corresponds to scenarios where replication is almost always the correct and cheapest choice. However, as $W_{BE}$ increases, the performance gap grows to over 200%. This is because a high $W_{BE}$ value creates a wide range of k-mer lengths for which synthesis is the cheaper option. The greedy algorithm falls into an "economic trap" by repeatedly choosing to pay the high fixed cost of $C_{reuse}$ for short k-mers, while the DP planner correctly identifies that synthesizing these fragments is the more cost-effective global strategy.

Conversely, the Max-Block Greedy algorithm, designed to minimize join operations by always selecting the largest possible k-mer ($W = 200$), exhibits the opposite behavior (**Figure 1a**). It performs poorly at low $W_{BE}$ values, showing a performance gap of over 100%. In this regime, the cost of synthesis is high, and the algorithm's forced choice to synthesize a large 200-mer (when a replicable version is unavailable) is a costly error compared to the DP planner's ability to find smaller, cheaper, replicable blocks. However, as $W_{BE}$ increases, the cost of synthesis becomes negligible. The Max-Block strategy of minimizing $C_{join}$ costs aligns with the DP algorithm's optimal strategy, and its performance gap rapidly drops to nearly zero. These results illustrate our central finding: while simple greedy heuristics can perform well under specific and limited economic conditions, only a provably optimal algorithm like our DP planner is robust enough to guarantee a cost-effective genome construction plan across all possible economic scenarios.

**Impact of source genome similarity on construction cost**

To investigate the relationship between the optimal construction cost and the evolutionary distance of the source genome, we conducted a computational experiment using the SARS-CoV-2 reference genome as a fixed target. A curated set of source genomes was selected, primarily from the Coronaviridae family, to provide a range of genetic similarities. The similarity between each source and the target was quantified using the Average Nucleotide Identity (ANI), a robust measure of genome-wide sequence identity, which we estimated using the MinHash algorithm implemented in the MASH software (Ondov et al. 2016). We then calculated the minimum construction cost under three distinct economic scenarios: "Replication-Dominant," "High-Join-Cost," and "Balanced.".

The results demonstrate a strong negative correlation between the construction cost and the ANI of the source genome across all tested scenarios (**Figure 1b-d**). As the source genome becomes more genetically similar to the target, the normalized cost per base decreases dramatically. This trend is most pronounced in the "Replication-Dominant" scenario, where the cost plummets by over 85% as the ANI increases from 36.79% to 100%. This occurs because higher similarity allows the Dynamic Programming planner to utilize a greater number of large, cost-effective $C_{reuse}$ operations, significantly reducing the reliance on expensive $C_{synth}$ operations. The data point at 36.79% ANI represents the baseline construction cost when using a completely dissimilar source (Enterobacteria phage EcoDS1, Enterobacteria phage phiX174), where the Mash distance is maximal. In this case, no significant k-mer reuse is possible, and the cost converges to a maximum determined almost entirely by de novo synthesis, effectively representing the cost of building the target genome from scratch. These results demonstrate that our cost-minimization framework capitalizes on genetic conservation and that the economic feasibility of synthetic genome construction is intrinsically linked to the presence of a closely related template organism.

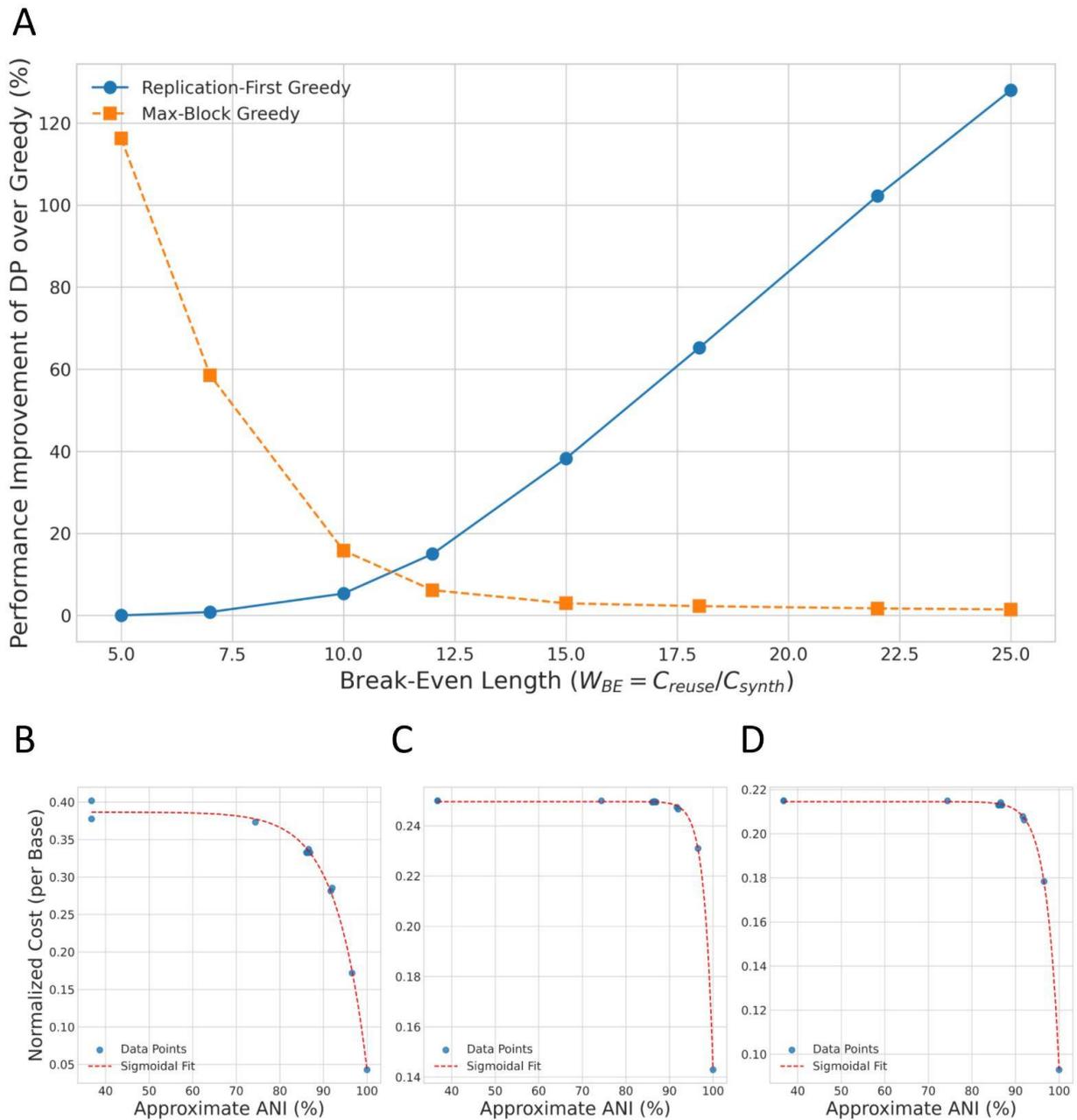

**Figure 1: Algorithmic Performance vs. Source Genome Similarity. (A)** The performance gap between the optimal DP planner and two greedy heuristics as a function of the Break-Even Length. **(B, C, D)** The relationship between the optimal construction cost and the source genome similarity (ANI %) under three distinct cost regimes: **(B)** Replication-Dominant, **(C)** High-Join-Cost, and **(D)** Balanced.

## Cost signatures distinguish conserved and rapidly evolving genomic regions

To investigate the relationship between our economic cost model and the functional architecture of the genome, we generated a "cost profile" by calculating the optimal construction cost over a sliding window for the SARS-CoV-2 target genome. This analysis was performed first using a single, closely related source genome (Bat CoV RaTG13) and subsequently averaged across multiple diverse coronavirus sources to obtain a consensus view.

| Gene | Single Profile Mean Cost | Grouped Analysis Mean Cost | Grouped Analysis Std Dev |
|---|---|---|---|
| ORF1ab | 0.1346 | 0.2082 | 0.0134 |
| Spike (S) | 0.1647 | 0.2124 | 0.0055 |
| ORF3a | 0.1441 | 0.2104 | 0.0110 |
| Envelope (E) | 0.0979 | 0.1934 | 0.0253 |
| Membrane (M) | 0.1175 | 0.1966 | 0.0216 |
| Nucleocapsid (N) | 0.1166 | 0.1986 | 0.0194 |

**Table1:** Combined Mean Cost Analysis per Gene Region.

The results, summarized in the table, reveal a distinct cost signature that correlates strongly with known biological functions. In both the single-source and multi-source analyses, the Spike (S) gene region consistently incurs the highest mean construction cost. This finding is significant, as the Spike protein is a surface glycoprotein responsible for host cell receptor binding and is known to be a primary target of the host immune system. Consequently, it is under intense selective pressure to mutate, leading to high sequence divergence even among closely related viruses. Our framework captures this evolutionary volatility as a high economic cost, as the significant number of differing base pairs necessitates more frequent and expensive de novo synthesis operations.

Conversely, the large ORF1ab region, which encodes the conserved viral replication and transcription machinery, exhibits a significantly lower construction cost in the single-source comparison. This reflects its high degree of sequence conservation, which allows the planner to utilize numerous cost-effective replication operations. Interestingly, while the Spike region has the highest average cost in the multi-source analysis, its standard deviation is low, suggesting that its high degree of divergence from SARS-CoV-2 is a consistent feature across the coronavirus family. In contrast, other accessory genes like the Envelope (E) show a higher variance, indicating more diverse evolutionary patterns within the group.

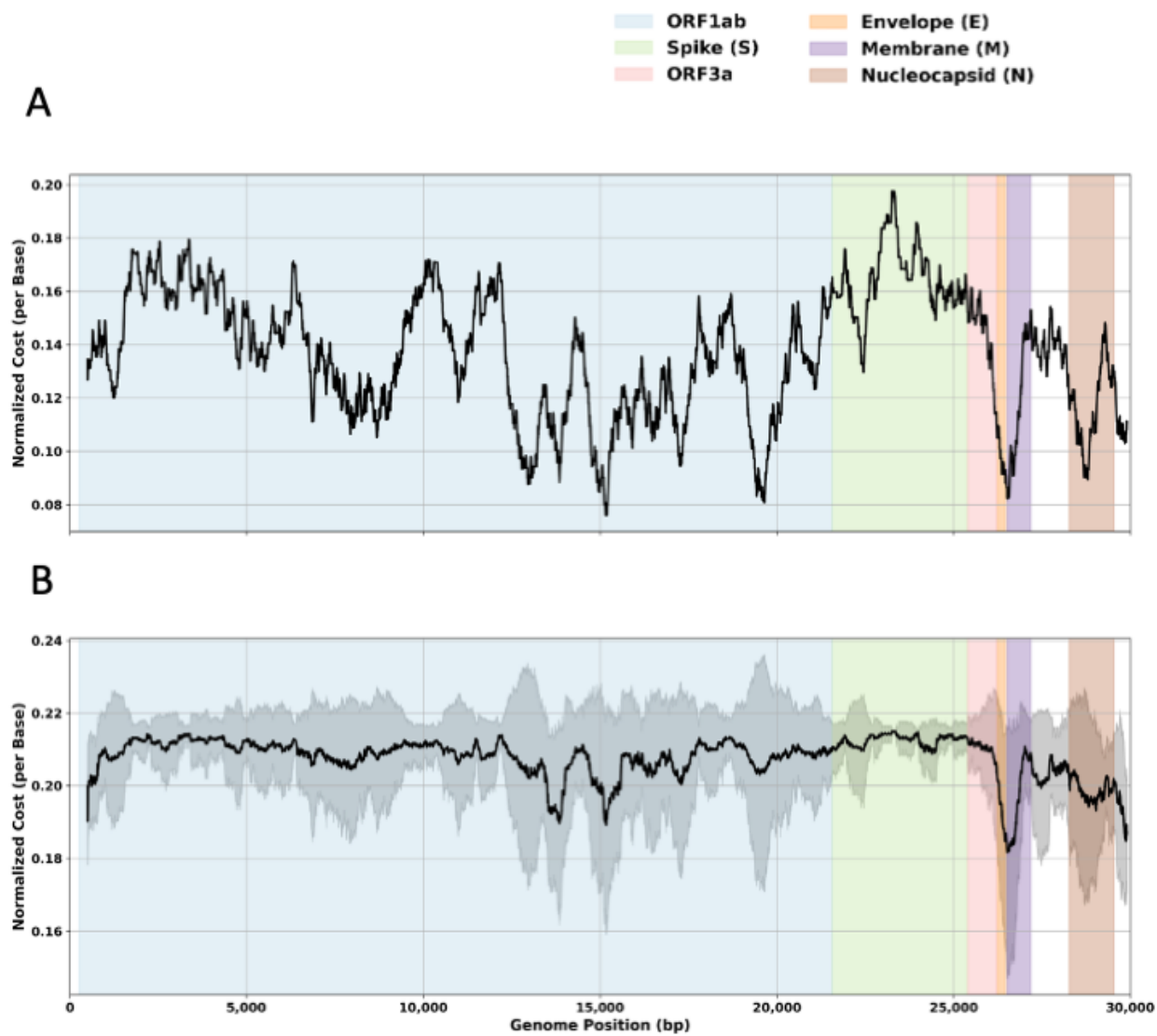

**Figure 2: Genome Architecture Cost Profile of SARS-CoV-2.** The optimal construction cost to build SARS-CoV-2 from **(A)** a closely related source (Bat CoV RaTG13) and **(B)** the mean cost from eight different coronavirus sources.

## Discussion

Our goal in this work was to model genome construction strategies that balance *de novo* DNA synthesis with the reuse of existing DNA fragments, thereby minimizing overall costs while enabling the efficient generation of long genomic sequences. Our study demonstrates that while greedy heuristics can perform well under specific cost regimes, the dynamic programming algorithm consistently identifies globally optimal solutions to the Minimum-Cost Synthetic Genome Planning problem across all scenarios. The strong correlation between construction cost and source-target genome similarity highlights that economic feasibility is tightly linked to evolutionary conservation, with closely related genomes enabling dramatic cost reductions. Furthermore, gene-level cost profiling revealed that fast-evolving, highly variable regions, such as the S (spike) gene in the SARS-coV-2 genome, impose disproportionately high synthesis demands, whereas conserved regions like ORF1ab are more cost-efficient to reconstruct. Together, these results establish our framework as a principled tool for connecting algorithmic planning with evolutionary dynamics, enabling more realistic and cost-effective strategies for synthetic genome engineering.

**Code Availability**

All source code, analysis scripts, and data required to reproduce the findings of this study are open-source and publicly available in two separate repositories. The standalone C++ implementation of the dynamic programming algorithm is available at GenomePlanner: https://github.com/Georgakopoulos-Soares-lab/GenomePlanner. This command-line tool is designed for efficiency, leveraging the sdsl-lite library (Gog et al. 2014) to construct an in-memory FM-index of the source genome for rapid substring queries.

The tool takes as input a source and a target genome in FASTA format and the key cost parameters from our model: --W for maximum block length (W), --pcr for the reuse cost ($C_{reuse}$), --join for the join cost ($C_{join}$), and --synth for the per-base synthesis cost ($c_s$).

To ensure the full reproducibility of our findings, all analysis pipelines, processing scripts, and configuration files used to generate the figures and results in this study are available at Minimum_cost_Genome_Planner: https://github.com/Georgakopoulos-Soares-lab/Minimum_cost_Genome_Planner.

**Acknowledgements**

Research reported in this publication was supported by the National Institute of General Medical Sciences of the National Institutes of Health under award number R35GM155468 and start-up funds awarded to I.G.S.